# Optimal parking provision in multi-modal morning commute problem considering ride-sourcing service

**Qida SU***

School of Civil and Environmental Engineering, Nanyang Technological University, 50 Nanyang Avenue, Singapore 639798

## Abstract

Managing morning commute traffic through parking provision management has been well studied in the literature. However, most previous studies made the assumption that all road users require parking spaces at CBD area. However, in recent years, due to technological advancements and low market entry barrier, more and more e-dispatch FHVs (eFHVs) are provided in service. The rapidly growing eFHVs, on one hand, supply substantial trip services and complete the trips requiring no parking demand; on the other hand, imposes congestion effects to all road users. In this study, we investigate the multi-modal morning commute problem with bottleneck congestion and parking space constraints in the presence of ride-sourcing and transit service. Meanwhile, we derive the optimal number of parking spaces to best manage the commute traffic. One interesting finding is that, in the presence of ride-sourcing, excessive supply of parking spaces could incur higher system commute costs in the multi-modal case.

* Corresponding author. Tel: (65) 89418262

Email: qida.su@ntu.edu.sg

# 1. Introduction

Morning commute traffic congestion is one of major problems faced by many large cities in the world. Commuters choose travel mode, as well as departure time, to minimize their individual travel costs. Every weekday, morning commuters take autos or public transit to reach their workplaces. If they decide to take auto, they can select either to drive their own cars that requires parking spaces or simply take a For Hire Vehicles (FHV) with no need of parking. Through departure time choice, commuters aim to avoid traffic congestion, as well as secure a parking space if they drive their own cars.

Indeed, using parking provision for morning commute traffic management has received more and more attention in the literature. Qian et al. (2011b) found that the availability of parking spaces can be a significant factor in commuters' travel decision making. Different parking permit distribution schemes were later compared by Zhang et al. (2011) in a many to one bottleneck model with the paralleled transit line. Yang et al. (2013) further proposed that an appropriate proportion of reserved parking spots can temporally alleviate road congestion and consequently increase social benefit. Interestingly, most parking management studies have made a similar assumption that all road users require parking space at CBD area and ignored the non-parking road users constituted mostly by for-hire vehicles (taxis and limousines) or the emerging and rapidly growing ride-sourcing service vehicles.

In fact, in recent years, due to technological advances and low market entry barrier, more and more e-dispatch FHVs (eFHVs) are provided in service, e.g., Uber in the US, Didi in China and Grab in Singapore. Riders first make instant bookings via mobile-app or phone call, then trip requests are distributed to eFHV drivers nearby. Drivers pick up passengers at the appointed origin and send them to the destination. Rayle et al. (2014) termed such service as ride-sourcing service. Some studies have been carried out on ride sourcing service or related app-based e-hailing service in terms of its market equilibrium, (He & Shen, 2015; Zha et al., 2016; Qian & Ukkusuri, 2017; Xu et al., 2017) and pricing strategy (Wang et al., 2016; Zha et al., 2017). Moreover, comparisons are also made with the traditional street hailing taxi service (Nie, 2017) or ridesharing (Dias et al., 2017). In fact, in the presence of the e-dispatch technology, once-distinct regulatory categories of FHVs are now blurring (*For Hire Vehicle Transportation Study*, 2016). Unlike ridesharing, which has been widely discussed in previous studies (Qian & Michael Zhang, 2011a; Furuhata et al., 2013; Xiao et al., 2016), the drivers in ride sourcing autos do not necessarily share a proximate destination or schedule with their passengers. Such difference has altered their demand for parking. In other words, while ridesharing autos require parking, yet similar to street hail taxis, most of ride sourcing autos do not need parking after arriving at the destination. For convenience, in this paper, we hereinafter refer to those vehicles which provide FHVs-like service as "eFHV" (if there are no special instructions), including the traditional FHVs, the emerging e-dispatch ride-sourcing vehicles as well as the driverless FHVs.

Thus, given the rapidly increasing scale of eFHVs, the common assumption without considering the FHVs in previous parking management studies may not be reasonable anymore. The eFHVs, which share the same road network with private autos, could impose a substantial impact upon the traffic congestion as well as the parking policy-making during morning peak. On one hand, a portion of commuters, who wish to avoid the crowdedness in transit service or no longer be able to secure a parking space in the city center, would switch to eFHVs. This will leave the well-designed traffic management via parking as studied in previous studies less efficient. On the other hand, the high and intensive morning commute demand could attract non-commuters to work as a part-time eFHV drivers and hence impose significant traffic congestion effects to all road users. Thus, in this study, we assume that the fleet size of eFHVs is not negligible anymore as compared to the counterpart of private autos with parking demand, and would examine how the eFHV service provision would affect the parking management measures.

Specifically, this paper aims to examine how the consideration of the eFHVs would affect the management of morning commute traffic with parking space constraints. To the best of our knowledge, eFHVs' effects on parking management and thus morning commute traffic management have not been explicitly addressed in previous studies. In this study, the equilibrium commute traffic taking into account surging supply and demand of eFHVs is completely depicted. From parking management perspective, we address the question on how to use the parking provision for best managing morning traffic by obtaining the optimal number of parking space provision. From eFHV's regulation perspective, we answer the question on whether and how we should control the number of registered eFHVs if the objective of managing morning traffic is to be achieved.

The rest of the paper is organized as follows. Section 2 revisits the classical bi-modal morning commute problem with parking constraints firstly, then further expand it to a multi-modal problem. Section 3 studies the equilibrium commuting pattern. We then analyze the network performance and introduce the optimal number of parking spaces and eFHVs to improve traffic efficiency in Section 4. Finally, Section 5 wraps up the paper with some summaries and future research directions.

## 2. Model formulation

### *2.1 Basic setting*

Consider a traffic corridor connecting a CBD and residential area as shown in Figure.1. A highway with a bottleneck and a paralleling transit line are available to commuters. During morning peak, homogenous travelers with same value of time $\alpha$ have three options commuting from home to workplace: drive alone, take eFHVs or catch a train. While only transit commuters use railway, both private auto drivers and eFHV passengers share the highway together. Let $N^C$, $N^F$ and $N^T$ denote the number of private auto, eFHV and transit commuters respectively. The total number of commuters, $N = N^C + N^F + N^T$, is assumed to be given. Commuters have a common preferred arrival time $t^*$ at the destination, and early or late arrival is penalized (schedule penalty for a unit time of early or late arrival: $\beta$, $\gamma$). We assume $\gamma > \alpha > \beta > 0$ based on the empirical evidence (Small, 1982). It is also assumed that the CBD is abstracted as a point and all parking spots are located at this point so that walking time from parking spots to workplace is simply ignored. Both private autos and eFHVs are assumed to carry one commuter per vehicle per trip only and consideration of ride sharing would be covered in future studies.

### *2.2 Bi-modal equilibrium considering parking searching*

We first revisit the bi-modal morning commute bottleneck model in the absence of ride-sourcing ($N^F = 0$), wherein all commuters using private auto need a parking spot at the destination. Every morning, road users need to tradeoff between the travel time cost related to bottleneck queue length and the scheduling cost of being early or late at work. If there is a binding parking capacity, drivers who do not have reserved parking spot would have to compete with each other for parking with earlier departure, and thus bear an extra time cost resulted from parking space searching while those who have parking reservation can be free from such worry and enjoy a lower commute cost (Yang et al., 2013). The parking searching time for private auto commuters departing at time $t$ is assumed to be a decreasing function of the number of non-occupied parking spaces at destination at the time when he left the bottleneck, namely an increasing function of the number of non-reserved vehicles departing before $t$ (Axhausen et al., 1994). Thus from the classical morning commute ADL model (Arnott et al., 1990), private auto travel cost for commuters departing at time $t$, in a modified form including travel time cost, schedule delay cost and parking fee, can be defined as follows:

$$\begin{aligned} \mathrm{P}_c(t) = \alpha \cdot (T(t) + S(t)) + \beta \cdot \max\left\{0, t^* - t - T(t) - S(t)\right\} \\ + \gamma \cdot \max\left\{0, t + T(t) + S(t) - t^*\right\} + F, \end{aligned} \tag{1}$$

where $T(t)$ is the travel time for departure time $t$ on the highway prior to parking, $S(t)$ is the parking searching time for vehicle departing at time $t$, $F$ is the parking fee and it is assumed to be a one-off fee.

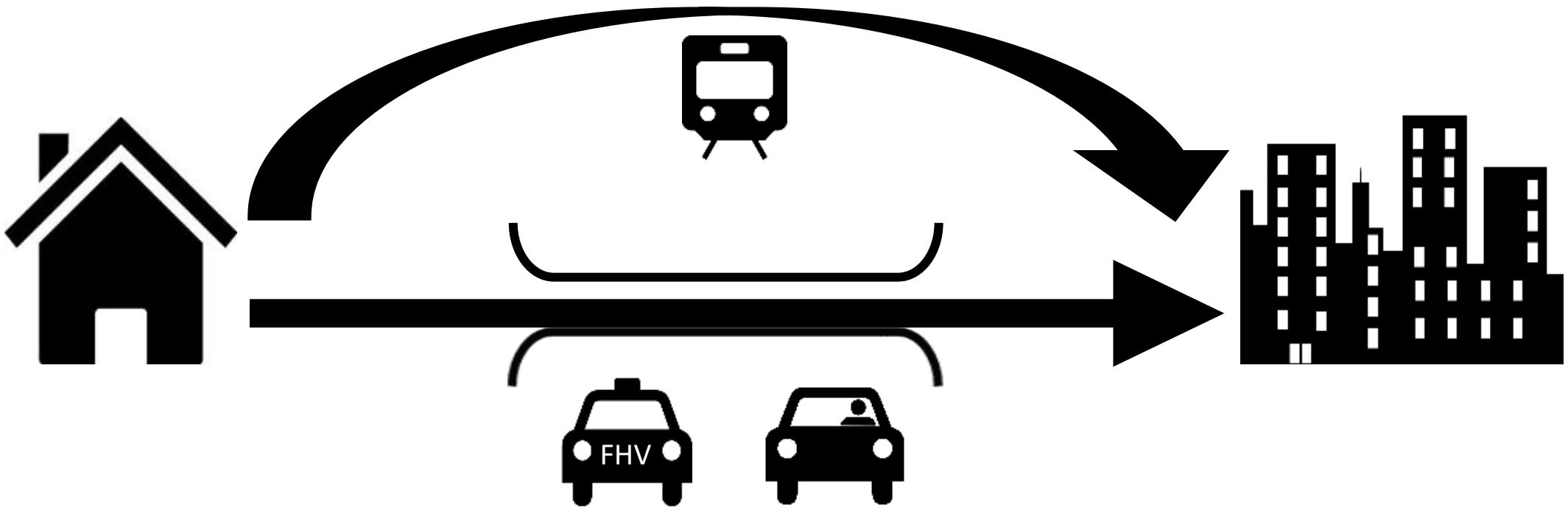

Figure.1. Morning commute corridor network

Here, free-flow travel time is assumed to be zero without loss of generality, and hence the highway only contains a bottleneck with a constant service capacity $s$ ($s > 0$), i.e. $T(t) = D(t)/s$, where $D(t)$ is the queue length at the bottleneck at $t$. In addition, the parking searching time is

$$S(t) = S_0 + \varepsilon \cdot \frac{\int_{t_1}^{t} r_c(u)\,du}{N^C}, \tag{2}$$

where $S_0$ is the initial searching time, $t_1$ is the earliest departure time from home for all road users, $r_c(t)$ is the flow rate of private auto entering the bottleneck at time $t$, $N^C$ is total number of private autos, or total number of parking spaces if there is binding parking capacity, and $\varepsilon$ is a constant coefficient that satisfies for private autos without parking reservation and $\varepsilon = 0$ for reserved private autos.

From within-day equilibrium, the arrival rates at the bottleneck for private auto commuters arriving at destination before and after $t^*$ respectively are given by

$$r_{c1} = \frac{s\alpha}{(\alpha - \beta)\left(1 + \dfrac{s\varepsilon}{N^C}\right)}, \quad r_{c2} = \frac{s\alpha}{(\alpha + \gamma)\left(1 + \dfrac{s\varepsilon}{N^C}\right)}. \tag{3}$$

The equilibrium commute cost from home to workplace for private auto drivers without parking restraint then become

$$p^C\left(N^C\right) = \frac{\beta\gamma}{\beta + \gamma}\frac{N^C}{s} + \alpha S_0 + \frac{\beta(\alpha + \gamma)}{\beta + \gamma}\varepsilon + F. \tag{4}$$

As for the transit commuters, we assume that the operating capacity of the railway is sufficiently large to the extent that the unpunctuality cost is not considered for transit commuters while the crowdedness in the transit still needs to be considered. Therefore, including the crowding effects, the commute cost of transit travelers $p^T$ can be given as an increasing function of the number of commuters using transit. As was done in Huang (2000), we let

$$p^T\left(N^T\right) = R + \theta \cdot N^T, \tag{5}$$

where $R$ and $\theta$ represent the transit fare and unit cost of crowding or discomfort from home to work on transit respectively.

From $p^C\left(N^C\right) = p^T\left(N^T\right)$ and $N = N^C + N^T$, we can obtain the virtual demand of parking space in the absence of ride-sourcing $N^{C0}$,

$$N^{C0} = \frac{s(\beta + \gamma)}{\mathbb{C}}\left[\theta N + R - \alpha S_0 - F + \frac{\beta(\alpha + \gamma)}{\beta + \gamma}\varepsilon\right], \tag{6}$$

where $\mathbb{C} = (\beta + \gamma)\theta s + \beta\gamma$, $N^{C0} \geq 0$. Obviously, $\mathbb{C} > 0$.

When there is binding parking capacity, i.e. $M < N^{C0}$, where $M$ is the number of parking spaces available at the destination ($M > 0$), the number of private auto commuters is restrained by $M$, and hence the number of transit commuters become $N - M$. In the absence of ride sourcing service, the equilibrium travel patterns of private auto commuters when there's parking constraint are similar to those shown in Figure 1 and 2 in (Yang et al., 2013), but with different departure rates and later arrival at the workplace due to the consideration of parking searching time in our setting. For simplicity and illustration purpose, we omit the sketches of these results here. Note that when all parking spaces are reserved, the arrival rates at the bottleneck for commuters arriving at destination before and after $t^*$ are given by $r_1 = \alpha s / (\alpha - \beta)$ and $r_2 = \alpha s / (\alpha + \gamma)$, where $r_1$ and $r_2$ are the special forms of $r_{c1}$ and $r_{c2}$ correspondingly when $\varepsilon = 0$ based on our setting for reserved parking vehicles. According to Eq. (4), the commute cost of private auto drivers in all-reserved case $P^{r0}$ is

$$P^{r0} = p^C(M) = \frac{\beta\gamma}{\beta + \gamma}\frac{M}{s} + \alpha S_0 + F \,. \tag{7}$$

*2.3 Multi-modal case*

Now we turn to the case in which the ride-sourcing service is considered ($N^F > 0$). Suffering from the parking space headache or transit over-crowdedness, some commuters are willing to choose other alternative transportation mode. The fast-developing ride sourcing service provides them such option. We postulate that all these eFHVs do not need parking at the destination at CBD [1] and only one home to work trip with one passenger can each eFHV driver completes during morning commute. After finishing the trip, they may travel around or leave the CBD without using the inbound capacity of highway. Additionally, the pricing mechanism of eFHVs for passengers is assumed to be same without surge pricing in this city regardless of the various ride-sourcing providers and depends only on the travel distance. Thus, in this model where all commuters travel along the same traffic corridor with constant distance, the fare for eFHV commuters are the same and constant. It is also assumed that commuters have no preference in choosing a specific eFHV and the waiting time for eFHV after booking is an equal constant for everyone independent of customers' booking time[2]. Every morning, eFHV commuters first make instant booking of eFHV service via mobile-app or phone call, then wait for the assigned eFHV to service. A commuter who make the booking earlier will board the eFHV and start the trip earlier as well. The time when a passenger boards his/her eFHV is assumed to be his/her departure time. Based on the assumptions above, the eFHV travel cost, departing at time $t$, is defined:

$$\mathrm{P}_f\left(t\right)=\alpha\cdot T\left(t\right)+\beta\cdot\max\left\{0,t^{*}-t-T\left(t\right)\right\}+\gamma\cdot\max\left\{0,t+T\left(t\right)-t^{*}\right\}+W\,, \tag{8}$$

where $W$ aggregates the generalized cost of eFHV other than travel time cost and schedule cost, which is principally composed of the travel fare and the waiting cost.

[1] Though some eFHVs are in fact ridesharing vehicles and drivers need to park as well after arriving at the destination, we focus on ride-sourcing FHVs in this study and thereby make this assumption. After all, drivers offer an eFHV ride in exchange for a fare income more often.(Rayle et al., 2014)

[2] We make this assumption for convenience of analysis. Some previous studies in the literature applied a matching function to determine the customers' average waiting time in traditional taxi analysis, depending on the customer demand and taxi utilization rate(Yang & Yang, 2011). As for ride sourcing service, when the number of eFHVs is sufficient, it is possible that the waiting time for ride-sourcing service becomes very short, with the aid of current advanced technology. In fact, the average waiting time of eFHV customers can be obtained in an aggregate model using the matching function given the current eFHV supply.

Since travel fare and waiting time are assumed to be constant, $W$ is a constant in our model. In comparison with the private auto travel cost, eFHV travel cost cuts out the parking-related cost, i.e., parking fee and parking searching time, and add the exclusive cost of eFHV instead.

When considering the ride sourcing service, the bottleneck services not only private auto commuters but also eFHV commuters at constant capacity and the bottleneck queue length $D(t)$ now become

$$D(t) = \max\left\{\int_{t_1}^{t}\left[r_c(u) + r_f(u)\right]\mathrm{d}u - s(t - t_1), 0\right\}, \tag{9}$$

where $r_f(t)$ is the flow rate of eFHVs entering the bottleneck at time $t$. At departure time $t$, if $\min\{\mathrm{P}_f(t), \mathrm{P}_c(t)\} > p^T(N^T)$, commuters without parking reservation will choose transit to work; conversely if $\min\{\mathrm{P}_f(t), \mathrm{P}_c(t)\} < p^T(N^T)$, they will use the road transport to workplace and will choose the mode (either private auto or eFHV) with the lowest travel cost at time $t$, if available. As for reserved parking commuters, they will drive to work only if their commute cost is equal to or lower than any other road users' travel cost. Furthermore, the number of available parking spaces is still $M$, and one can verify easily that $N^C \le M$. The actual number of private auto commuters $N^C$ is now not only bounded by the parking spaces $M$ but also by the eFHVs $N^F$. The performance of ride-sourcing service can alter commuters' mode choice. This can be termed as a deterministic Wardropian multi-modal equilibrium. Consequently, the commute cost of non-reserved private auto drivers (denoted by $P^u$), eFHV riders (denoted by $P^f$) and transit passengers, are identical at equilibrium, and higher than that of reserved commuters (denoted by $P^r$), i.e.,

$$P^u = P^f = p^T\left(N - N^C - N^F\right) \geq P^r \ . \ ^3 \tag{10}$$

Regarding within-day equilibrium, it should satisfy

$$\frac{\mathrm{d}}{\mathrm{d}t}P_c\left(t\right) = 0 \text{ and } \frac{\mathrm{d}}{\mathrm{d}t}P_f\left(t\right) = 0. \tag{11}$$

## 3. Multi-modal equilibrium considering ride-sourcing

In this section, we investigate the multi-modal morning traffic equilibrium on the condition that the supply of ride-sourcing service is not limited. Namely, we assume that there is sufficiently large fleet size of eFHVs to meet the commute demand.[4] Meanwhile, parking supply $M$ falls short of demand, and all parking spaces are not reserved.

Since the private auto drivers have to reserve some time for parking other than passing through the bottleneck, for the same preferred arrival time, the private auto driver arriving on time will always depart earlier than the on-time arrival eFHV passengers. Let $t_{3\mathrm{c}}$ and $t_{3\mathrm{f}}$ denote the departure time of on-time arrival private autos and eFHVs, and we have $t_{3\mathrm{c}} < t_{3\mathrm{f}}$. Thus, the morning rush hours can be divided into three time windows: Both private autos and eFHVs arrive early; private autos arrive late while eFHV arrive early; both private autos and eFHVs arrive late.

**Lemma 3.1** *There is no simultaneous departures of private autos and eFHVs at equilibrium at any* $t$.

---

[3] Here, it is reasonable to assume $P^u \geq P^r$. If $P^u < P^r$, it means that those who have parking reservation in advance have to pay even higher cost during morning commute compared to those who have not parking reservation. In the end, no one would like to make parking reservation and all private auto commuters will compete for parking.

[4] This seems to be impractical at first glance, however if the market entry barrier of being an eFHV drivers become low enough and the extra income from driving eFHV is attractive, or if the technology of autonomous car is well developed and driverless cars can be widely employed in FHV industry, the supply of eFHVs can possibly satisfy its demand.

From (11), if there is a time point when both private autos and eFHVs have departures simultaneously, the departure rates of private autos and eFHVs should satisfy

$$\begin{cases} \left(1+\frac{s\varepsilon}{N^C}\right)r_c + r_f = \frac{s\alpha}{(\alpha-\beta)} = r_c + r_f, & \text{when } t \le t_{3c}; \\ \left(1+\frac{s\varepsilon}{N^C}\right)r_c + r_f = \frac{s\alpha}{(\alpha+\gamma)}, r_c + r_f = \frac{s\alpha}{(\alpha-\beta)}, & \text{when } t_{3c} < t \le t_{3f}; \\ \left(1+\frac{s\varepsilon}{N^C}\right)r_c + r_f = \frac{s\alpha}{(\alpha+\gamma)} = r_c + r_f, & \text{when } t > t_{3f}; \end{cases} \tag{12}$$

where $r_c, r_f \ge 0$,

when both private autos and eFHVs have departures, $r_c, r_f > 0$. Obviously, for each time window, we cannot find such positive $r_c$ and $r_f$ respectively to satisfy the equilibrium conditions as shown above. In other words, it is impossible to have simultaneous positive departures of the two modes at equilibrium. Therefore, for private autos, the early and late arrival departure rates are $\mathrm{r}_{c1}$ and $\mathrm{r}_{c2}$, while the early and late arrival departure rates of eFHVs are $\mathrm{r}_1$ and $\mathrm{r}_2$ as long as there is positive departure. This is also consistent with the findings in Lindsey (2004).

Let $W_0 = (\alpha-\beta)\cdot S_0 + F$. When $W \le W_0$, from Eqs. (1) and (8), there always exists $\mathrm{P}_f(t) \le \mathrm{P}_c(t)$ for any $t$, which means that riding eFHVs is better than driving during morning peak. When eFHVs' supply is sufficient, all commuters will not drive to work, and only eFHVs use the highway.

When $W > W_0$, private auto driving is used for commuting. Since $\mathrm{P}_c(t_1) < \mathrm{P}_f(t_1)$, there are only private autos departures at $t_1$. However, the latter travel pattern of bottleneck becomes different depending on the parking spaces as well as other parameters. At first private autos arrive at the bottleneck at a constant rate $r_{c1}$ and no eFHV is on the road, the commute cost of eFHV riders gradually decrease and eventually equals to the cost of private auto drivers. The time point with equal commute cost may occur before or after the full occupancy of all parking spaces.

**Lemma 3.2** *Once the commute costs of the two modes equal, i.e., at* $t = t_2$ *, there is no longer any private autos departure when* $t > t_2$ *and the road will be filled with eFHVs only, if the supply of eFHV is sufficient.*

This lemma can be proved by contradiction. Suppose there exists a time point $t_2$, where $\mathrm{P}_f(t_2) = P$, the equilibrium commute cost. If there is still positive private auto departure from $t_2$ to $t_2 + \Delta t$ ($\Delta t > 0$), we have $\mathrm{P}_c(t_2) = \mathrm{P}_c(t_2 + \Delta t)$ and no eFHVs departures. When $t_2 \le t_{3\mathrm{c}}$, according to (1),

$$\mathrm{P}_c(t_2 + \Delta t) - \mathrm{P}_c(t_2) = (\alpha - \beta)\left[\frac{(r_{c1} - s)}{s}\Delta t + \frac{\varepsilon r_{c1}}{M}\Delta t\right] - \beta \Delta t = 0. \qquad (13)$$

The difference between the commute cost of eFHVs at $t_2 + \Delta t$ and $t$ is

$$\begin{aligned} &\mathrm{P}_f(t_2 + \Delta t) - \mathrm{P}_f(t_2) = (\alpha - \beta)\left[\frac{(r_{c1} - s)}{s}\Delta t\right] - \beta \Delta t < 0, \\ &\rightarrow \mathrm{P}_f(t_2 + \Delta t) < P. \end{aligned} \qquad (14)$$

Likewise, it can be proved that when $t_{3\mathrm{c}} < t_2 \le t_{3\mathrm{f}}$ or $t_2 > t_{3\mathrm{f}}$, $\rightarrow \mathrm{P}_f(t_2 + \Delta t) < P$.

We find that the commute cost of eFHVs at $t_2 + \Delta t$ will always be lower than the equilibrium commute cost and this contradicts the principle of user equilibrium. Thus, at equilibrium, no one will choose private autos anymore after $t_2$. If bottleneck queues have not been cleared at $t_2$, there must be eFHV departures at later times, and the last road user takes eFHVs to work. Otherwise, it is possible that no one chooses eFHVs when the cost of eFHV is pretty high or when eFHVs departing at $t_2$ arrives already late. On this account, we notice that the number of parking spaces may not always be binding in the case when $W > W_0$ since the private auto driving becomes no longer cost competitive after a specific time point.

**Lemma 3.3** *When both private autos and eFHVs are used for commuting, the total number of road users* $N^F + N^C$ *is a constant independent of* $N^C$ *or* $M$ *,*

$$N^0 = \frac{(\beta+\gamma)s}{\mathbb{C}}\left(R + \theta N - K\right)$$
$$\text{where } K = \frac{(\alpha-\beta)\gamma}{\beta+\gamma}S_0 + \frac{\beta}{\beta+\gamma}W + \frac{\gamma}{\beta+\gamma}F, \tag{15}$$

*if the bottleneck queue has never been cleared before the last arrival of road users, i.e., when bottleneck services at full capacity without a break during morning commute.*

If there are both private autos' and eFHVs' departures, we have

$$N^F + M \geq N^F + N^C = \left(t_4 - t_1\right)s, \tag{16}$$

$$\mathrm{P}_c\left(t_1\right) = \mathrm{P}_f\left(t_4\right), \tag{17}$$

where $t_4$ denotes the last departure time of road users during morning peak. Let $M_2^u = \frac{\left[W - \left(F + (\alpha-\beta)S_0\right)\right]s}{\beta}$, then after some substitutions, we have

$$t_4 = \frac{(\beta+\gamma)t^* - \beta t_1 - \beta M_2^u / s}{\gamma}, \tag{18}$$

$$
\begin{aligned}
&N^F + M \geq N^F + N^C = \frac{(\beta+\gamma)}{\gamma}\left(t^* - t_1\right)s - \frac{\beta}{\gamma}M_2^u, \\
&\quad \to t^* - t_1 = \frac{\gamma\left(N^F + N^C\right) + \beta M_2^u}{(\beta+\gamma)s}, \\
&\quad \to P = \mathrm{P}_c\left(t_1\right) = \frac{\beta\gamma}{(\beta+\gamma)s}\left(N^F + N^C\right) + K = p^T\left(N - N^F - N^C\right), \\
&\quad \to N^F + N^C = \frac{(\beta+\gamma)s}{\mathbb{C}}\left(R + \theta N - K\right) = N^0.
\end{aligned}
$$

Therefore, the individual equilibrium commute cost does not change after $N^C$ and thereby does not change after $M$ when there does not ever exist an empty bottleneck before the end of morning peak.

Now we attempt to determine the travel pattern of road users. Let

$$
W_1 = (\alpha-\beta)(S_0+\varepsilon)+F, \quad M_1^u = \frac{p^T\left(N-N^0\right) - F - \alpha S_0 - \beta\varepsilon}{\beta}s,
$$

$$
W_2 = (\alpha+\gamma)(S_0+\varepsilon)+F, \quad M_3^u = \frac{p^T\left(N-N^0\right) - W^0}{\beta}s,
$$

$$
M_4^u = \frac{(\alpha+\gamma)\varepsilon}{W-W^2}M_u^3,
$$

$$
M_5^u = \frac{(\beta+\gamma)[p^T\left(N-N^0\right) - F - \alpha S_0] - \beta\left(W-W^0\right) - (\alpha+\gamma)\beta\varepsilon}{\beta(\beta+\gamma)}s.
$$

Apparently, $W_0 < W_1 < W_3$, $M_1^u < M_3^u$, $M_2^u < M_3^u$. Then the domains of different commuting travel patterns with respect to total number of parking spaces $M$ and the generalized cost of eFHVs other than traveling and scheduling cost $W$, is shown in Figure 2. Note that we only consider the morning commute problem with parking spaces restraint, i.e., $M < N^{C0}$.

Figure 3 (a) - (j) then depict the possible equilibrium commuting patterns, where blue curve AB and H(F)E represent the cumulative arrival at the bottleneck of private autos and eFHVs respectively and red curve DC and GE represent their cumulative arrival at the destination respectively.[5] $t_{cl}$ represents the departure time of the last private auto during morning peak.

Note that in case (a) (c) (e) (f) (g) (h), both private autos and eFHVs are used, and the bottleneck queue has never been cleared before the last arrival of road users. From Lemma 3.3, the total number of road users in these cases equals to $N^0$ and the equilibrium commute cost $P_{(a)(c)(e)(f)(g)(h)} = p^T\left(N - N^0\right)$, which is independent of the number of parking spaces $M$.

Though the equilibrium cost does not change after $M$, all parking spaces are taken up in cases (c) (e) (f). Yet in cases (a) (g) (h), the actual number of private auto drivers

---

[5] Figure 3 (a) sketches the case when the commute cost of eFHVs and private autos become equal before the full occupancy of all parking spaces and both last private car and first eFHV arrive at CBD early. Figure 3 (b) and (c) sketch the cases when the commute cost of eFHVs and private autos become equal after the full occupancy of all parking spaces and both last private car and first eFHV arrive at CBD early. Figure 3 (d) (e) sketch the cases when the commute cost of eFHVs and private autos become equal after the full occupancy of all parking spaces, last private car arrives at CBD late while first eFHV arrives early. Figure 3 (f) sketch the case when the commute cost of eFHVs and private autos become equal after the full occupancy of all parking spaces and both last private car and first eFHV arrive at CBD late. Figure 3 (g) sketches the case when the commute cost of eFHVs and private autos become equal before the full occupancy of all parking spaces and last private car arrives at CBD late while first eFHV arrives early. Figure 3 (h) sketches the case when the commute cost of eFHVs and private autos become equal before the full occupancy of all parking spaces and both last private car and first eFHV arrive at CBD late. Figure 3 (i) sketches an extreme case when no one uses eFHV and commuters drive to work only during morning peak. Figure 3 (j) sketches another extreme case when no one uses car driving and commuters take eFHVs to work only during morning peak.

is even less than the number of parking spaces, which indicates the oversupply of parking spaces.

Now we compare and contrast the equilibrium with or without ride-sourcing service. According to Eq. (10), given the parking spaces $M$, the system commute cost difference (denoted by $\Delta$, the system commute cost with ride sourcing minus that without ride sourcing) $\Delta = \left[ p^T\left(N - N^0\right) - p^T\left(N - M\right)\right] \cdot N$. When $W \le W_2$, there is $N^{C0} \le N^0$. Since $M < N^{C0}$, $M < N^0 \to N - M > N - N^0 \to \Delta < 0$. When $W > W_3$, $N^{C0} > N^0$, whether $\Delta \le 0$ depends on the value of $M$. It cannot be determined directly whether the addition of ride-sourcing service into bottleneck problem will make the bottleneck performance at equilibrium better with lower system commute cost. Yet in most cases when there is under-supply of parking spaces, $M$ can be considered smaller than $N^0$, $\Delta < 0$, i.e., the introduction of ride-sourcing improves system performance.

As for case (b) and (d), both private autos and eFHVs are used and the first eFHV departs after the bottleneck queue is cleared. All parking spaces are used up during morning peak. The equilibrium commute cost become

$$P_{(\mathrm{b}),(\mathrm{d})} = \frac{\beta\gamma}{\mathbb{C}}\left[R + \theta\left(N - M\right) - W\right] + W = p^T\left(N - M - N^F\right).$$

It can be verified that the commute cost of road users in case (b) and (d) is a strictly decreasing function of $M$. As shown in Figure 2, the number of parking spaces should satisfy $M \le M_2^u$. At $M = M_2^u$, the equilibrium commute cost reaches the minimum $P_{(\mathrm{b}),(\mathrm{d})\min} = \frac{\beta\gamma}{\left(\beta+\gamma\right)s} N^0 + \mathrm{K} = p^T\left(N - N^0\right)$, namely $P_{(\mathrm{b}),(\mathrm{d})} \ge p^T\left(N - N^0\right)$. The system commute cost difference is computed as $\Delta = p^T\left(N - M - N^F\right) - p^T\left(N - M\right)$. Since $N^F > 0$, we always have $\Delta < 0$ in case (b) and (d).

In case (i), the number of road users equals to the number of parking spaces $M$. The equilibrium commute cost $P_{(\mathrm{i})} = p^T\left(N - M\right)$ is a strictly decreasing function of $M$. In comparison with the case without ride-sourcing, the system commute cost difference $\Delta = p^T\left(N - M\right) - p^T\left(N - M\right) = 0$.

Lastly in case (j), all commuters will not drive to work, and apparently the number of road users is independent of the number of parking spaces $M$. The equilibrium commute cost becomes $P_{(\mathrm{j})} = P^f = \frac{\beta\gamma}{\left(\beta+\gamma\right)}\frac{N^{F0}}{s} + W = p^T\left(N - N^{F0}\right)$, where $N^{F0} = \frac{s\left(\beta+\gamma\right)\left(R - W + \theta N\right)}{\mathbb{C}}$, representing the virtual demand of eFHVs.

Without loss of generality, we have $N^{F0} > 0$; otherwise no one uses the highway, and hence we can assume that $W < R + \theta N$ always exists. The system commute cost difference $\Delta = \left[p^T\left(N - N^{F0}\right) - p^T\left(N - M\right)\right] \cdot N$. When $N^{C0} < N^{F0}$ or $N^{C0} \geq N^{F0}$ and $M < N^{F0}$, one can derive that $\Delta < 0$; when $N^{C0} \geq N^{F0}$ and $M \geq N^{F0}$, we have $\Delta \geq 0$. In most cases when the parking spaces are not sufficient, we have $M < N^{F0}$, i.e., $\Delta < 0$.

From above analysis, one can find, among all the cases shown in Figure 5, only in case (b) (d) and (i), increasing the parking spaces supply can reduce the equilibrium commute cost. In other cases, the equilibrium commute cost is independent of the number of parking spaces when eFHVs are sufficient. However, the reasons behind lower commute cost in (b) (d) and (i) are different. In case (i), the ride-sourcing service is no longer attractive, and it is indeed the bi-model equilibrium. While in (b) (d), it is because the shortage of parking spaces is extremely acute, and the marginal benefit of parking spaces increment to system cost reduction remains high. Once the scale of parking spaces is expanded to a threshold, i.e. $M_2^u$, a further increase in parking facilities provision brings no benefit to the system performance. Therefore, the morning commute management via parking spaces control becomes ineffectual in many instances. Total number of road users can no longer be constricted by parking spaces provision as in Zhang et al. (2011), neither can the total queuing delay at bottleneck (a proxy of bottleneck performance). Yet the existence of ride-sourcing service can lower the system commute cost of commuters when there are insufficient of parking spaces, as it fills the unsaturated road capacity resulted from parking space paucity. Such system commute cost reduction induced by ride-sourcing also implies that the commute cost derived from the traditional bi-modal equilibrium is usually overestimated.

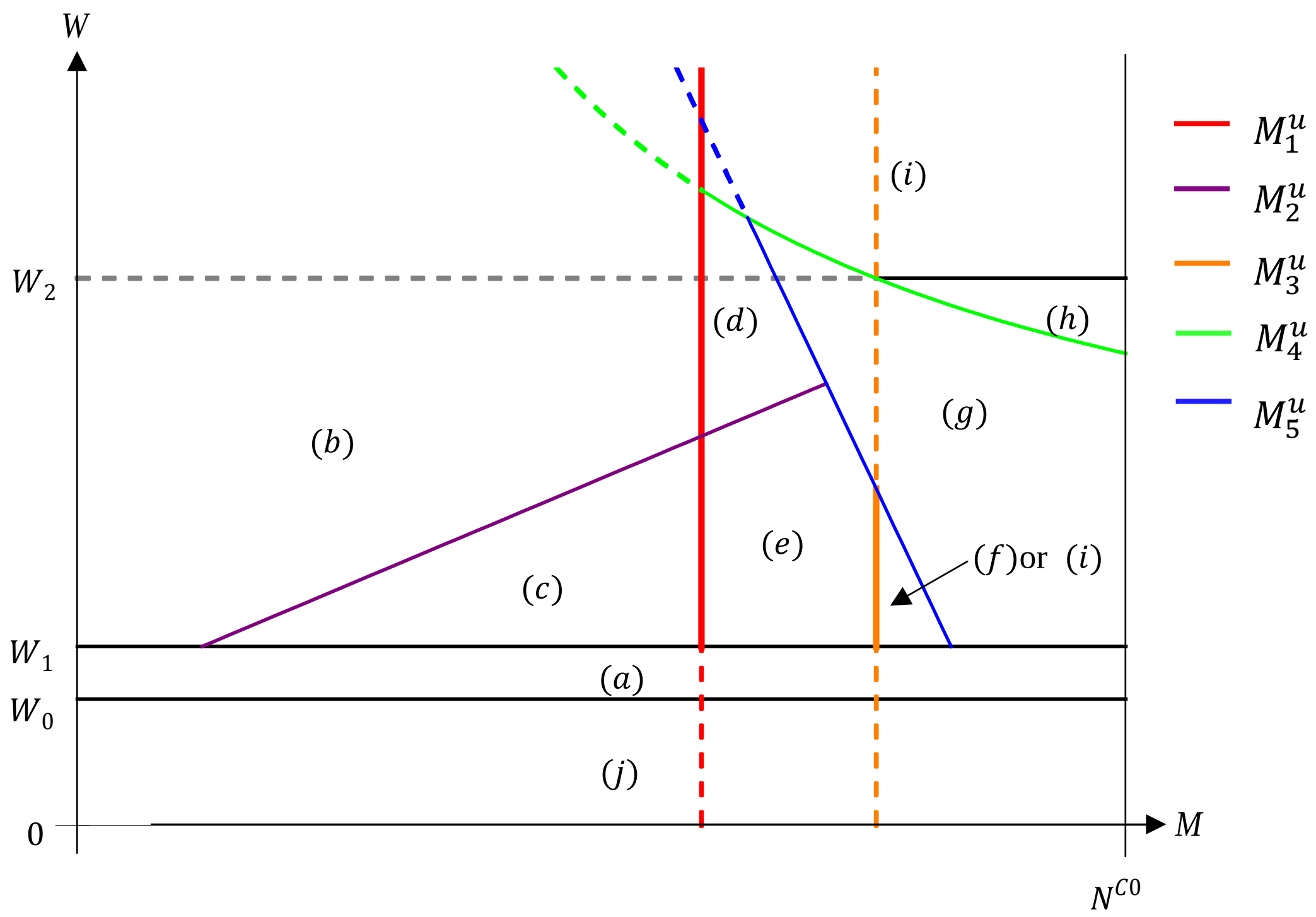

Figure 2: Domain of possible commuting patterns with parking space restraint in the presence of sufficient ride-sourcing service with respect to $M$ and $W$.

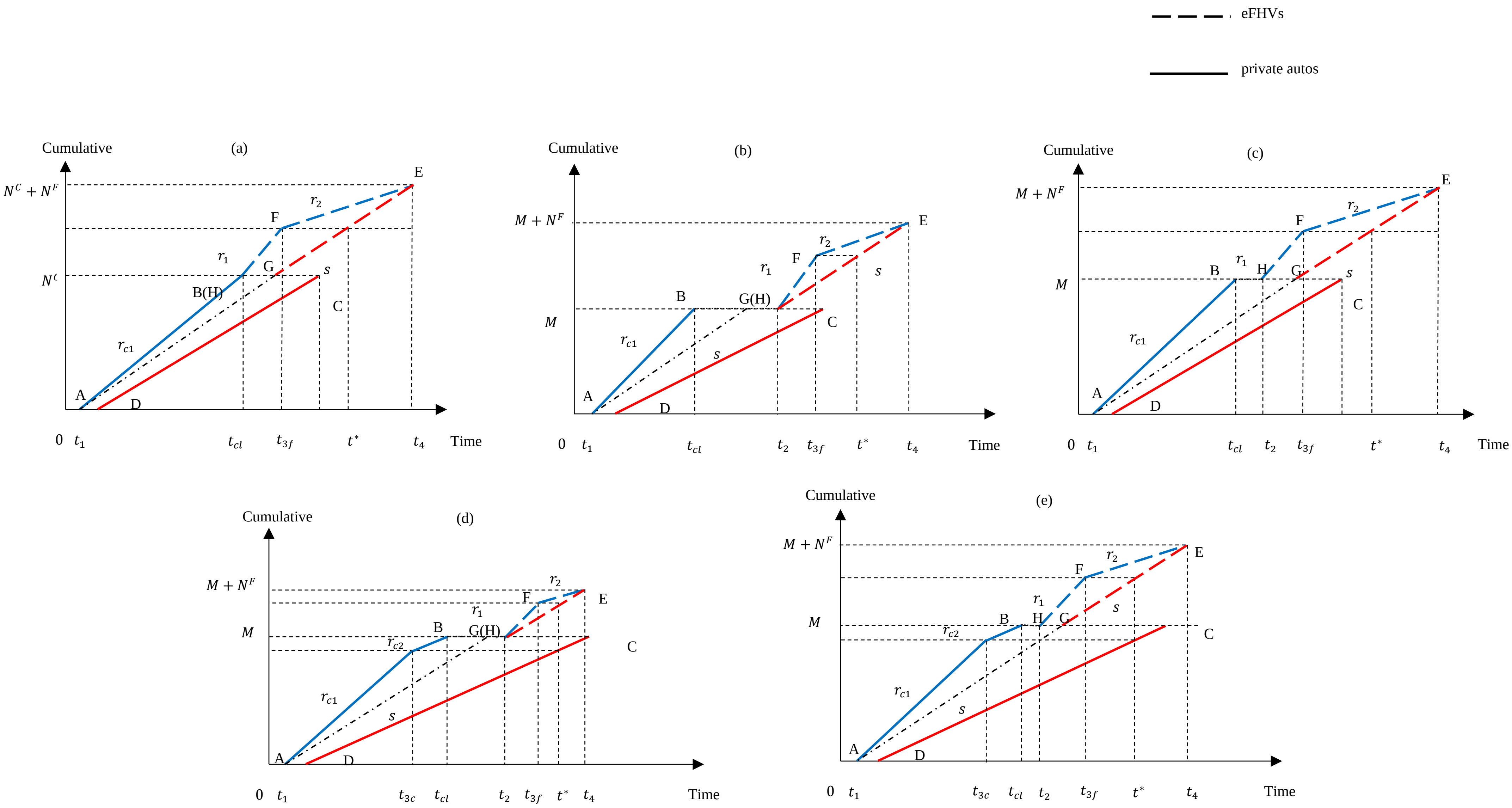

eFHVs
private autos
Cumulative
(a)
$N^C + N^F$
$N^C$
E
F
G
B(H)
A
D
C
$r_2$
$r_1$
$r_{c1}$
$s$
0
$t_1$
$t_{cl}$
$t_{3f}$
$t^*$
$t_4$
Time
Cumulative
(b)
$M + N^F$
$M$
E
F
B
G(H)
A
D
C
$r_2$
$r_1$
$r_{c1}$
$s$
0
$t_1$
$t_{cl}$
$t_2$
$t_{3f}$
$t^*$
$t_4$
Time
Cumulative
(c)
$M + N^F$
$M$
E
F
B
H
G
A
D
C
$r_2$
$r_1$
$r_{c1}$
$s$
0
$t_1$
$t_{cl}$
$t_2$
$t_{3f}$
$t^*$
$t_4$
Time
Cumulative
(d)
$M + N^F$
$M$
$r_2$
F
E
$r_1$
B
G(H)
C
$r_{c2}$
$r_{c1}$
$s$
A
D
0
$t_1$
$t_{3c}$
$t_{cl}$
$t_2$
$t_{3f}$
$t^*$
$t_4$
Time
Cumulative
(e)
$M + N^F$
E
$r_2$
F
$r_1$
$M$
B
H
G
$s$
C
$r_{c2}$
$r_{c1}$
$s$
A
D
0
$t_1$
$t_{3c}$
$t_{cl}$
$t_2$
$t_{3f}$
$t^*$
$t_4$
Time

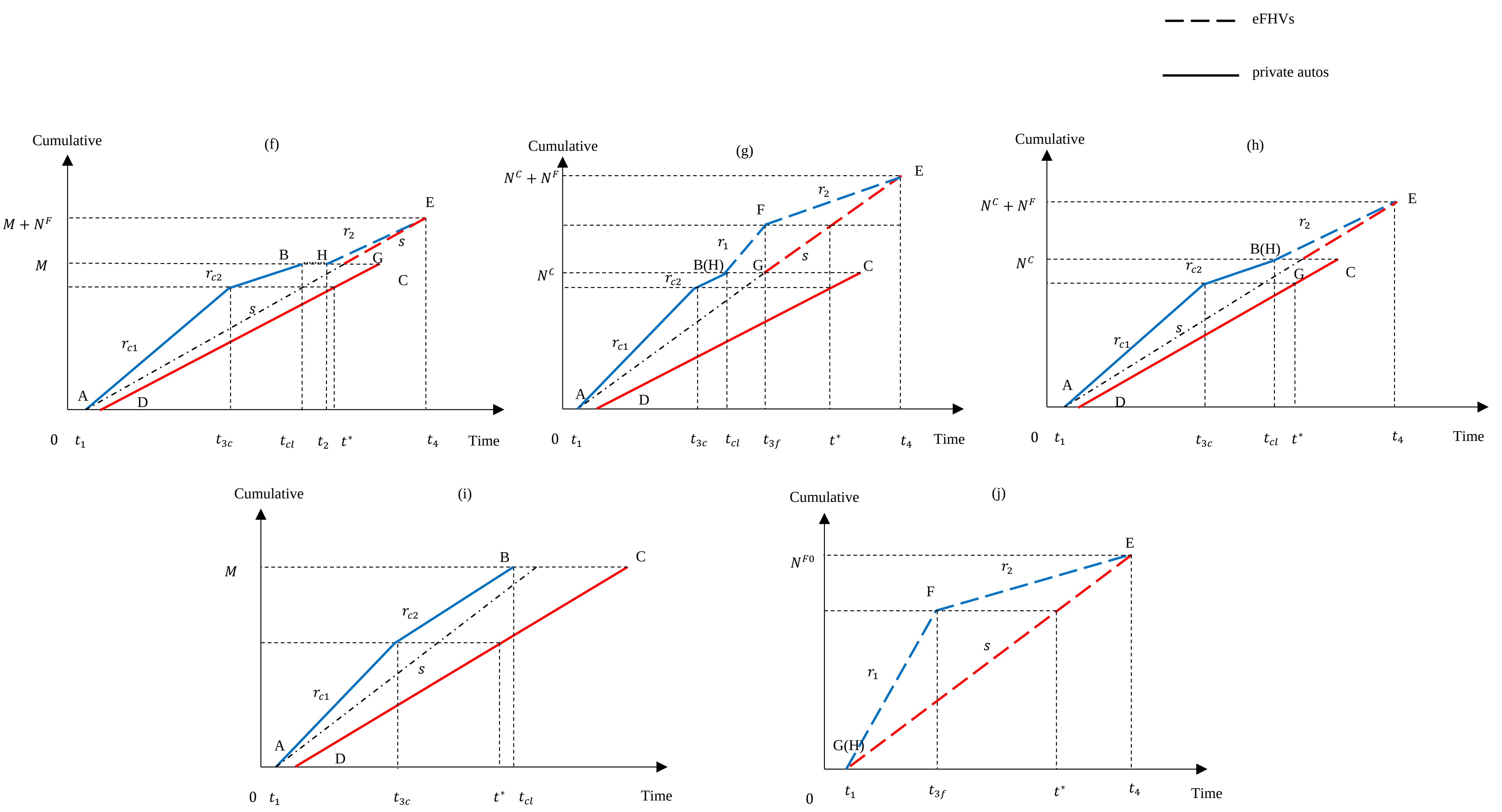

Figure 3 (a)-(j): Possible Commuting patterns with parking space restraint in the presence of sufficient ride-sourcing service

## 4. The optimal number of parking space and eFHVs in morning commute

Having solved the equilibrium morning traffic pattern considering ride-sourcing, we seek to address the problem that how we can obtain the optimal number of parking spaces from the perspective of best managing the morning commute traffic.

Likewise, parking spaces provided $M$ are assumed to be insufficient in this section. The number of eFHVs which can service morning commute, $N^F$, is no longer assumed to be sufficient and is smaller than the virtual demand $N^{F0}$.[6] Additionally, we assume $W \leq W_0$. Prior researches (Zhang et al., 2011; Yang et al., 2013) have suggested that system commute cost will always be reduced when all parking permits are distributed in advance, i.e., all parking spaces are reserved, compared to the case in which all parking spaces are not reserved. To this end, all private auto drivers in this part are assumed to have parking reservation[7], namely $\varepsilon = 0$ for all private autos.

Besides, several variables $\mathbb{D}$, $\mathbb{E}$, $\mathbb{A}$, $\mathbb{B}$, $N^{\mathrm{F1}}$ to $N^{\mathrm{F6}}$ and $M^1$ to $M^7$, are hereinafter defined for reference and comparison purpose. Let

[6] Various factors can affect the supply of eFHVs in practice, for instance, the income of the ride-sourcing drivers, the profit of ride-sourcing firms, the average demand of ride-sourcing service and the government policy. Zha et al. (2016) discussed the equilibrium supply of ride-sourcing vehicles in an aggregate model, which is not detailed in this study as it is beyond the focus of this paper. Once $N^F$ is given, the major issue addressed hereinafter is whether parking space control is still feasible in morning commute management, when eFHV is presented.

[7] When the parking spaces at CBD are all employer-provided or with monthly/yearly booking and no public parking are set aside, i.e. drivers are not allowed to park at CBD during morning peak without parking permit, we can state that all parking spaces have been reserved.

$$\mathbb{D}=W_0-W,\qquad N^{\mathrm{F1}}=\frac{\mathbb{D}+\mathbb{E}-\theta N}{2\beta+\theta \mathrm{s}}s,\qquad M^1=\frac{\beta+\gamma}{2}\frac{\mathbb{A}}{\mathbb{C}},$$

$$\mathbb{E}=\theta N+R-W-\beta S_0,\qquad N^{\mathrm{F2}}=\frac{\beta\left(\mathbb{D}+\mathbb{E}-\theta N\right)+\gamma\mathbb{E}}{2\beta^2+\mathbb{C}},\qquad M^2=\left(\beta+\gamma\right)\frac{\mathbb{B}}{\mathbb{C}},$$

$$\mathbb{A}=\left[\mathbb{E}-\mathbb{D}+\theta\left(N-N^F\right)\right]s,\qquad N^{\mathrm{F3}}=\frac{\mathbb{D}}{\beta}s,\qquad M^3=\frac{\beta\mathbb{A}+\gamma\mathbb{B}}{2\mathbb{C}},$$

$$\mathbb{B}=\left(\mathbb{E}-\theta N^F\right)s-\beta N^F;\qquad N^{\mathrm{F4}}=\frac{\mathbb{E}-\mathbb{D}}{\theta},\qquad M^4=\left(\beta+\gamma\right)\frac{\mathbb{A}-\theta Ns}{\mathbb{C}},$$

$$N^{\mathrm{F5}}=\frac{\beta\left(\mathbb{E}-\mathbb{D}\right)+\gamma\mathbb{E}}{\mathbb{C}}s,\qquad M^5=N^{\mathrm{F5}}-N^{\mathrm{F}},$$

$$N^{\mathrm{F6}}=N^{\mathrm{F5}}-M^6,\qquad M^6=\frac{\beta\theta Ns}{\mathbb{C}},$$

$$N^{\mathrm{F7}}=\frac{\mathbb{E}s-\dfrac{\mathbb{C}}{\beta+\gamma}M}{\beta+\theta \mathrm{s}},\qquad M^7=N^{\mathrm{F5}}-N^{\mathrm{F3}}.$$

$$N^{\mathrm{F8}}=N^{\mathrm{F4}}-\frac{\mathbb{C}}{(\beta+\gamma)\theta \mathrm{s}}M,$$

$$N^{\mathrm{F9}}=N^{\mathrm{F5}}-M;$$

Obviously, $\mathbb{D}\geq 0$ and $\mathbb{C}=\left(\beta+\gamma\right)\theta s+\beta\gamma$. From (6), the equilibrium private auto drivers without parking restraint $N^{C0}\geq 0$. When all drivers have parking reservation ($\varepsilon=0$), $\theta N+R-\alpha S_0-F\geq 0$.

Then we have $\mathbb{E}=\theta N+R-W-\beta S_0=\theta N+R-\alpha S_0-F+\mathbb{D}\geq 0$. Besides, one may verify that the equations $M=M^2$ and $N^{\mathrm{F}}=N^{\mathrm{F7}}$ are equivalent. Such equivalence also applies to: 1. $M=M^4$ and $N^{\mathrm{F}}=N^{\mathrm{F8}}$, and 2. $M=M^5$ and $N^{\mathrm{F}}=N^{\mathrm{F9}}$.

Following the setting above, the number of eFHV commuters and private auto commuters are limited to $N^F$ and $M$, respectively. Since ride-sourcing is more attractive than driving at any time point ($\mathrm{P}_f\left(t\right)\leq \mathrm{P}_c\left(t\right)$ for any $t$) in our setting, road users will compete for the insufficient eFHVs first with earlier trip booking and earlier departure. However, as shown in Eq. (10), the equilibrium commute cost for eFHV passengers is less than that of private auto drivers with parking reservation. This is because private auto mode will not be chosen until full occupancy of all eFHVs and eFHVs commuters have to spend more on scheduling due to such eFHV competition. As a result, all private auto drivers in this case depart later than eFHV passengers with a departure gap. The larger scheduling cost of eFHV passengers has finally turned eFHV's superiority over driving into inferiority.

Possible scenarios of cumulative departures from home and arrivals at destination at equilibrium depending on different $N^F$ and $M$ are presented below.[8] The commute cost for each scenario including schedule delay and queuing delay costs is then calculated. Time points for each scenario are presented in Table 1.

Scenario 1): All eFHV passengers arrives at destination early (before $t^*$). The first private auto departs after the bottleneck queue is cleared and hence the bottleneck services without flow interaction between eFHVs and private autos.

Scenario 2): All eFHV passengers arrives at destination early. Some private autos enter the bottleneck before the queue is cleared and wait for pass till all eFHVs leaves the bottleneck.

Figure 4 depicts the critical case between Scenario 1) and Scenario 2) in which the first private auto happens to enter the bottleneck right after the last eFHV leaves the bottleneck. Curve AB and GF represent the cumulative arrival at the bottleneck of private autos and eFHVs respectively and curve DC and GE represent their cumulative arrival at the destination correspondingly.

[8] There is no such scenario when some eFHV passengers arrive at destination late and some private autos arrive early. Note $t_{3f} > t_{3c}$ as discussed in Section 3. In this case, we should have the last eFHV departure time $t_{fl} > t_{3f}$, the first private auto departure time $t_2 < t_{3f}$ and hence $t_2 < t_{fl}$, contradicting the fact that all private autos depart later than eFHVs.

Scenario 3): Some eFHV passengers arrive at destination late (after $t^*$). All private autos arrive late. From (11), it is impossible to have late-arrival departure after the queue is cleared. So the first private auto can only depart before the bottleneck queue is cleared.

The individual travel cost and the total travel cost of all commuters in the previous scenarios will be discussed in the following subsections: first to determine the optimal number of parking space $M^{\mathrm{O}}$ with a given $N^{\mathrm{F}}$ in 4.1 , and second to determine the optimal number of vehicles providing ride sourcing service $N^{\mathrm{FO}}$ with a given $M$ in 4.2 .

**Table 1 Time points for each scenario.**

| *Time* | Scenario 1) | Scenario 2) | Scenario 3) |
|---|---|---|---|
| $t_1$ | $t^* - \frac{P^f - W}{\beta}$ | | |
| $t_{fl}$ | $t^* - \frac{P^f - W}{\beta} + \frac{N^F}{s}$ | $t^* - \frac{P^f - W}{\beta} + \frac{\alpha-\beta}{\alpha}\frac{N^F}{s}$ | $t^* - \frac{\alpha+\beta+\gamma}{\alpha}\frac{P^f - W}{\beta} + \frac{\alpha+\gamma}{\alpha}\frac{N^F}{s}$ |
| $t_2$ | $t^* + \frac{P^r - (\alpha+\gamma)S_0 - F}{\gamma} - \frac{N^C}{s}$ | $t^* + \frac{\alpha}{\alpha+\beta+\gamma}\frac{P^r - (\alpha+\gamma)S_0 - F}{\gamma} - \frac{\alpha+\gamma}{\alpha+\beta+\gamma}\frac{N^C}{s} - S_0$ | $t^* + \frac{P^r - (\alpha+\gamma)S_0 - F}{\gamma} - \frac{\alpha+\gamma}{\alpha}\frac{N^C}{s}$ |
| $t_4$ | $t^* + \frac{P^r - (\alpha+\gamma)S_0 - F}{\gamma}$ | | |

*4.1 Optimal number of parking space*

4.1.1 Analysis results

We first assume the number of eFHVs $N^F$ is given during morning commute. Though there are several scenarios upon different departure and arrival times, the individual commute cost for eFHVs riders and transit passengers are equal, i.e.

$P^f = p^T\left(N - N^F - N^C\right)$, according to Eq.(10). As for private auto drivers, their individual commute cost can be summarized in two cases:

$$P^r = \begin{cases} \dfrac{\beta\gamma}{(\beta+\gamma)s} N^C + \alpha S_0 + F, & \text{if } N^F \le N^{F3} \text{ or } N^F > N^{F3} \text{ and } M \le M^2 \text{ (i);} \\ \dfrac{\gamma\left[Ws + \beta\left(N^C + N^F\right) - P^f s\right]}{\beta s} + (\alpha+\gamma) S_0 + F, & \text{if } N^F > N^{F3} \text{ and } M > M^2 \text{ (ii);} \end{cases} \tag{19}$$

where $N^C \le M$ and if $N^F \le N^{F3}$, $N^C = \min\left\{M, M^4\right\}$; if $N^F > N^{F3}$, $N^C = \min\{M, M^5\}$.

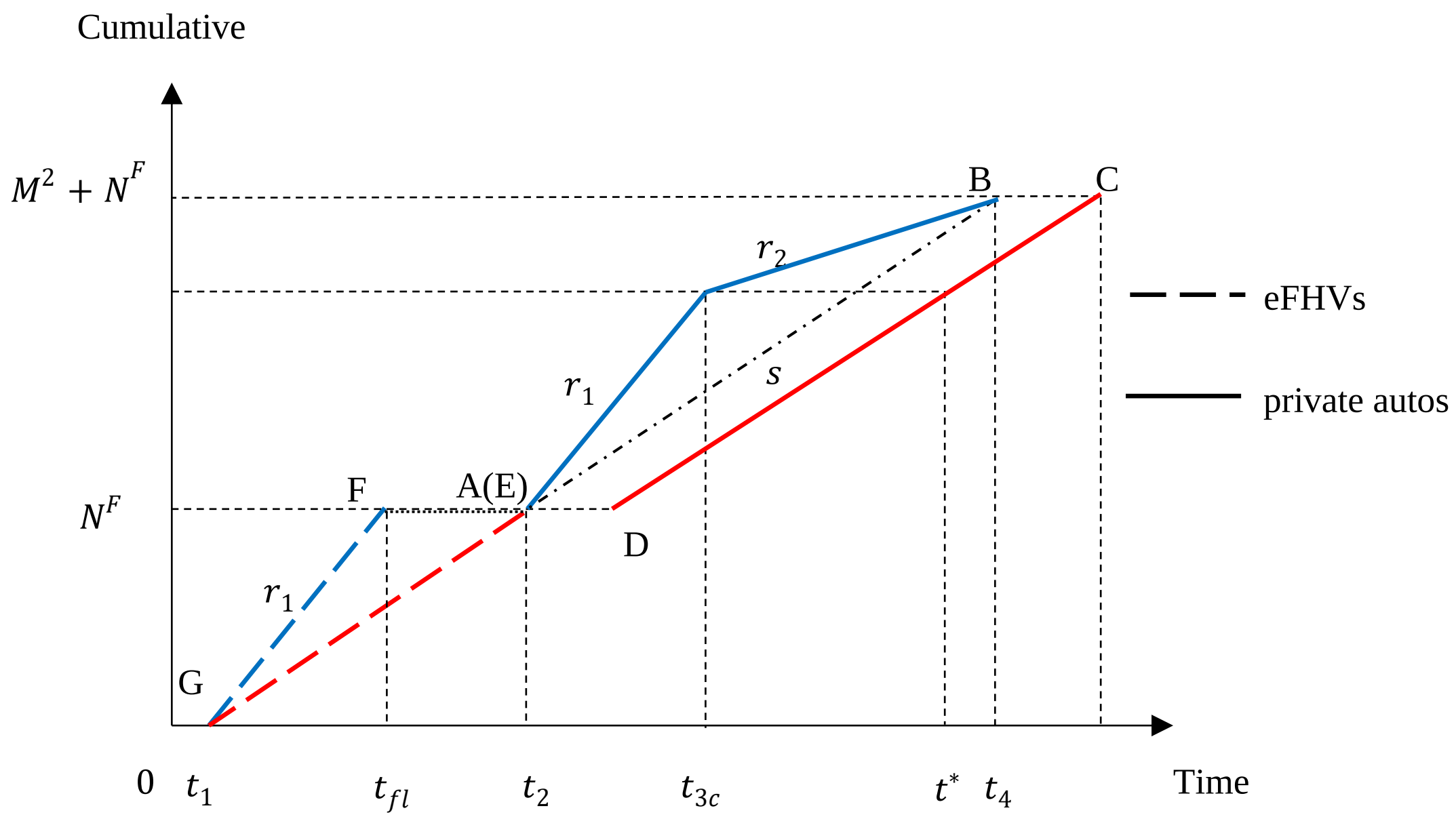

Figure 4: Critical commuting equilibrium between Scenario 1) and Scenario 2), with parking space restraint in the presence of insufficient ride-sourcing service, when $W \le W_0$.

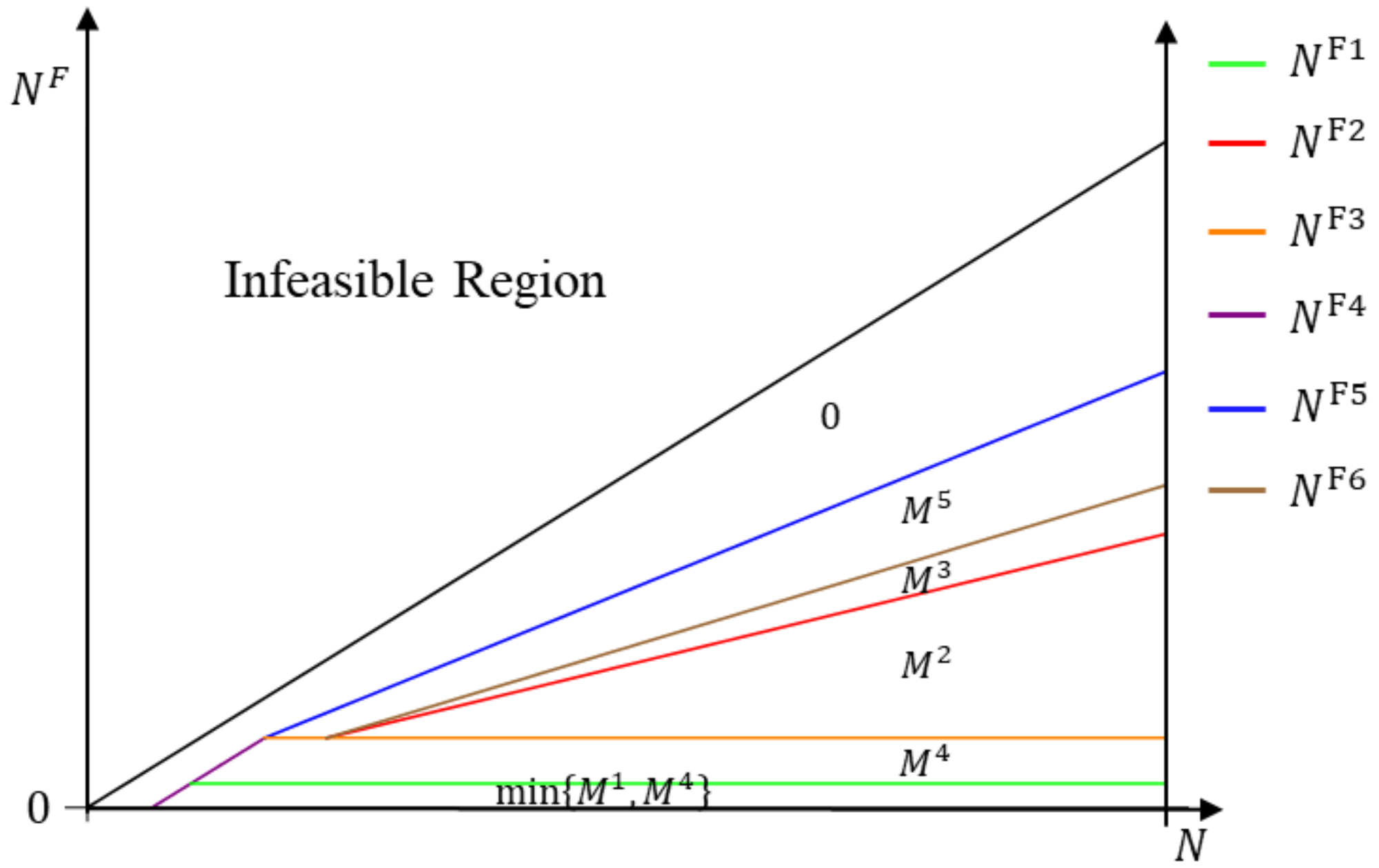

Figure 5: Domain of optimal number of parking spaces $M^O$ in the presence of insufficient ride-sourcing service with respect to $N$ and $N^F$, when $W \leq W_0$.

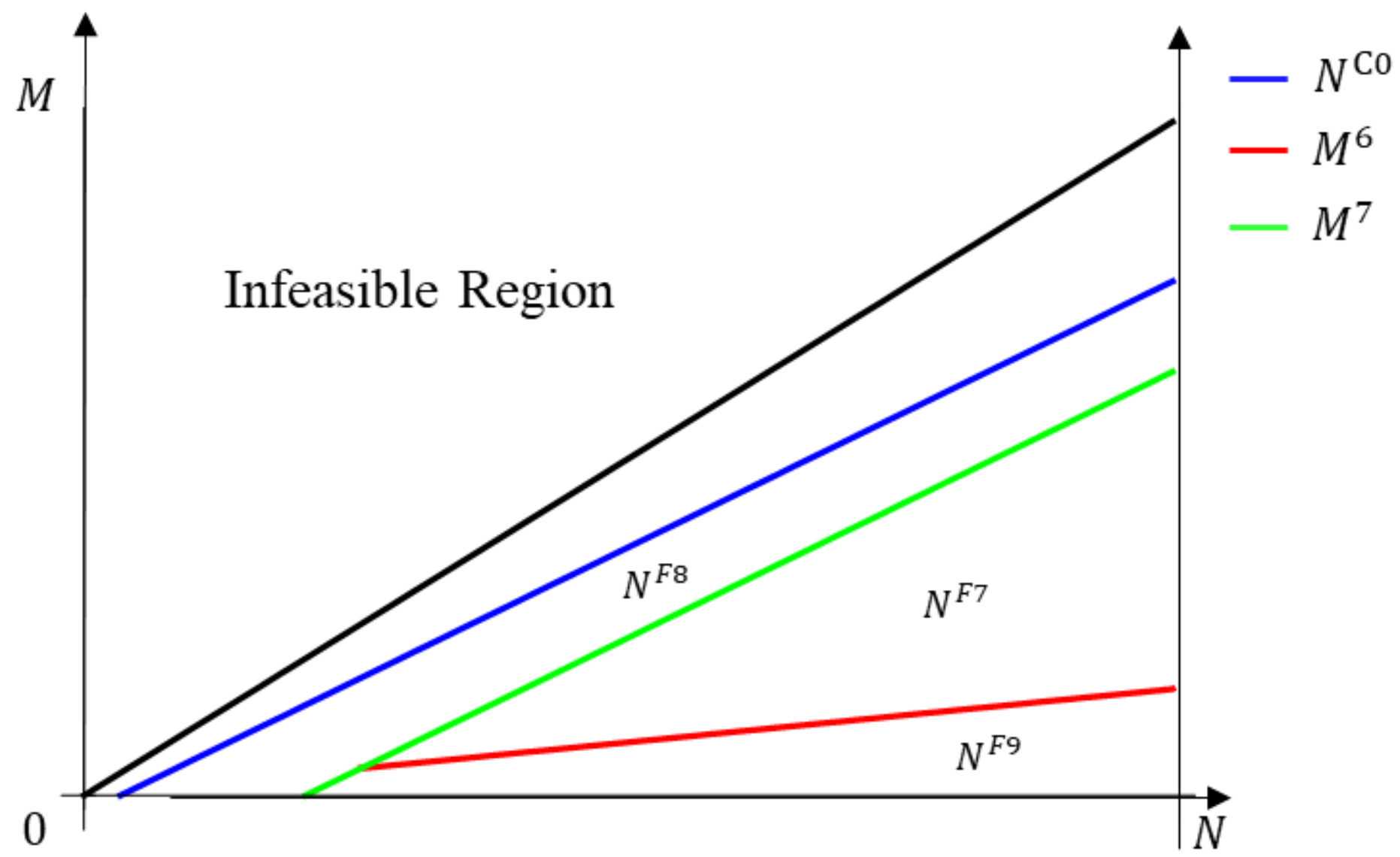

Figure 6: Domain of optimal number of eFHVs $N^{FO}$ with parking space restraint, with respect to $N$ and $M$, when $W \leq W_0$.

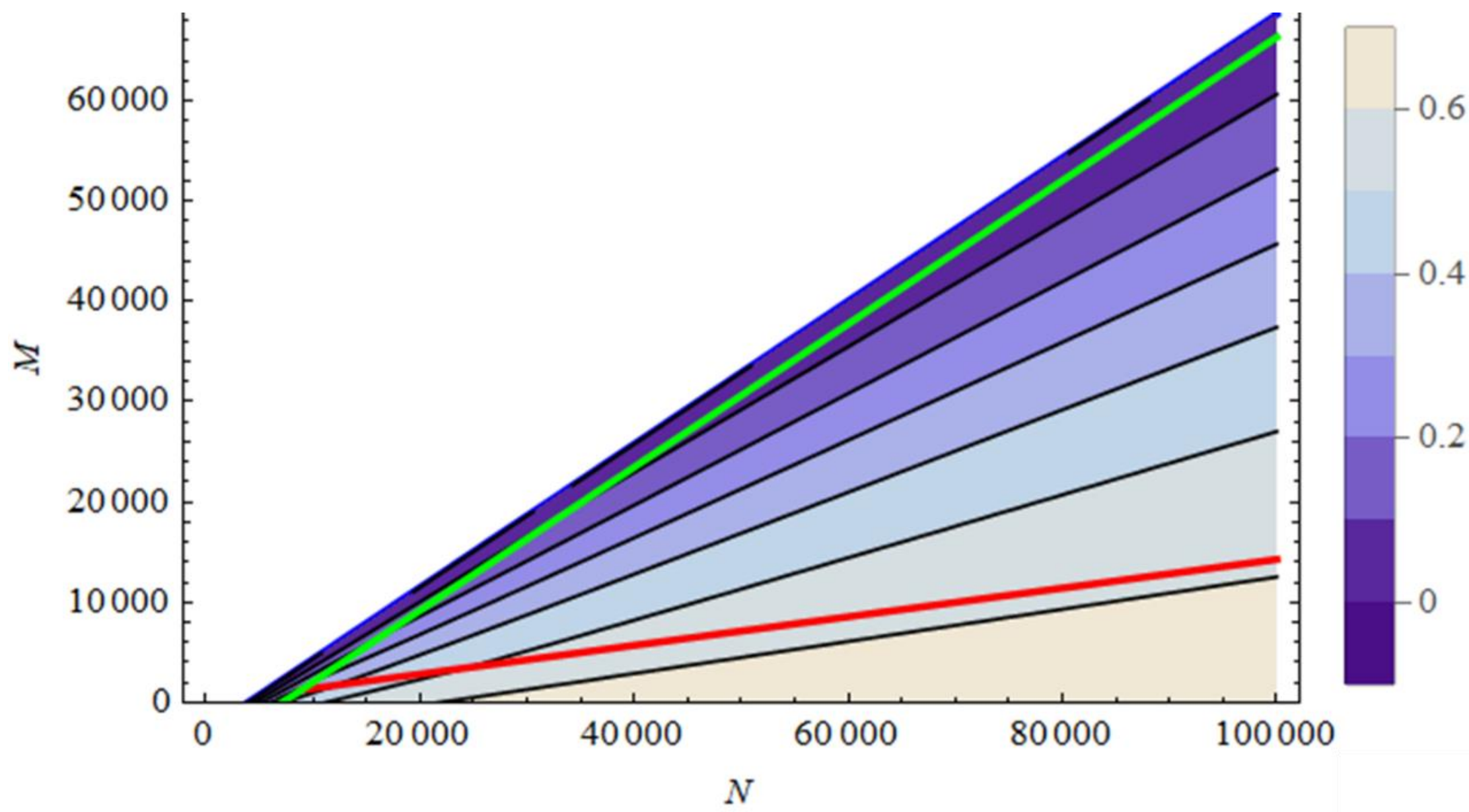

Figure 7: Efficiency contours $\eta$ with respect to $N$ and $M$, when $W \le W_0$.

Therefore, the parking spaces provided $M$ is binding only when $M$ is smaller than a specific value ($M^4$ or $M^5$ in this study); otherwise, the number of private auto commuters is less than $M$ and the parking space provision is redundant.

More intuitively, case (i) in Eq. (19) indicates that there exists one period with zero flow at the bottleneck, i.e., Scenario 1); case (ii) represents that there never exists a period with zero flow at the bottleneck, i.e., Scenario 2) and 3).

Now let's move on to the system commute cost (denoted by $TC$). $TC$ is the total cost for all commuters, including private auto cost, transit cost, and eFHV cost and is given as follows,

$$\begin{aligned} TC &= P^r \cdot N^C + P^f \cdot N^F + p^T\left(N - N^F - N^C\right)\cdot\left(N - N^C - N^F\right) \\ &= P^r \cdot N^C + P^f \cdot \left(N - N^C\right), \end{aligned} \tag{20}$$

where $P^r$ is determined in Eq.(19) and $P^f = p^T\left(N - N^F - N^C\right)$.

Given the number of eFHVs during morning peak $N^F$, as shown in Figure 5, we have the following proposition.

**Proposition 4.1** *For given* $N^F$*, there is always an optimal number of parking spaces* $M = M^O\left(N^F\right)$*, which leads the system commute cost to minimum.*

$$M^O(N^F)=\begin{cases}\min\{M^1,M^4\}, & if\ N^F \le N^{F4}\ and\ N^F \le N^{F1};\\ M^4, & if\ N^F \le N^{F4}\ and\ N^F > N^{F1}\ and\ N^F \le N^{F3};\\ M^2, & if\ N^F \le N^{F4}\ and\ N^F > N^{F3}\ and\ N^F \le N^{F2};\\ M^3, & if\ N^F \le N^{F6}\ and\ N^F > N^{F3}\ and\ N^F > N^{F2};\\ M^5, & if\ N^F \le N^{F5}\ and\ N^F > N^{F6}\ and\ N^F > N^{F3};\\ 0, & otherwise.\end{cases} \quad (21)$$

**Proof.** See Appendix A. 1. □

From Proposition 4.1, one can easily find that, with different levels of ride-sourcing service supply, the corresponding optimal parking spaces supply can be different. A mismatch parking supply results in social loss. When the current supply of parking spaces falls below the optimum at current $N^F$, system commute cost increases; conversely when the current supply of parking spaces outstrips this optimum, the system commute cost either increases or remains the same with unnecessary operating expense on redundant parking spaces. In terms of morning commute management, when the ride-sourcing service generally performs better than the private auto driving during morning peak ($W \le W_0$), the increase of parking spaces supply does not surely reduce system commute cost, considering ride-sourcing service. Instead, an optimal number of parking permits coordinated with the current number of eFHVs can improve travel efficiency. When necessary, it is cost effective to temporally block up the superfluous parking spaces (e.g., curbside parking) and limit the distributed parking permits to optimum during morning peak to achieve minimum system commute costs.

#### 4.1.2 Numerical example

We now use a numerical example to illustrate the results above. Following Jia et al. (2016), the value of time, schedule delay penalties are taken as follows: $\alpha = 0.3$ (SGD$/min), $\beta = 0.1$ (SGD$/min), $\gamma = 0.4$ (SGD$/min). The highway bottleneck capacity is $s = 200$ (veh/min). We further suppose $\theta = 0.001$ , $R = 1$ (SGD$/commuter), $F = 4$ (SGD$/commuter), $W = 3$ (SGD$/commuter), $S_0 = 3$ (min). Then for instance, when the total number of commuters is $100,000$ , if parking provision meets the virtual parking demand $68,643$, we have system commute cost $TC = 3.24\times 10^6$ (SGD$) without ride-sourcing service. Now with $10,000$ eFHVs, the total cost becomes $TC = 3.14\times 10^6$ (SGD$), and the actual number of occupied parking space is $59,557$ with $9,086$ parking space idle. By reducing the available parking space to $59,071$ for morning commute, we can further lower the system commute cost to $TC = 2.99\times 10^6$ (SGD$).

### *4.2 Optimal number of eFHVs*

In previous subsection, the optimal parking provision is obtained. However, one may argue that the number of parking space is difficult to be manipulated. Particularly for those office buildings, once the construction is finished, the provided parking spaces can be hardly changed. In this subsection, we then focus on the question from the eFHVs regulation perspective, i.e., what is the optimal number of eFHVs, when the number of parking space $M$ is given, so that the system travel cost can be minimized. Indeed, it is imperative for the regulators to address the question that: should the number of registered eFHVs be controlled? If yes, how the fleet size of eFHVs should be determined? Granted that the answer to such a question is stipulated by many different factors and considerations, we can at least attempt to address this problem in this study from the perspective of managing morning commute traffic.

#### 4.2.1 Analysis results

The individual commute cost of private auto users is summarized below:

$$P^r = \begin{cases} \dfrac{\beta\gamma}{(\beta+\gamma)s}N^{\mathrm{C}} + \alpha S_0 + F, & \text{if } N^F \le N^{F7} \text{ or } N^F > N^{F7} \text{ and } M \ge M^7 \text{ (i);} \\ \dfrac{\gamma\left[Ws + \beta\left(N^{\mathrm{C}} + N^F\right) - P^f s\right]}{\beta s} + (\alpha+\gamma)S_0 + F, & \text{if } N^F > N^{F7} \text{ and } M < M^7 \text{ (ii);} \end{cases}$$

where $N^{\mathrm{C}} \le M$ and if $M \ge M^7, N^{\mathrm{C}} = \min\{M, M^4\}$; if $M < M^7, N^{\mathrm{C}} = \min\{M, M^5\}$.

(22)

The individual cost of private auto users is almost the same as that in 4.1 , except the slightly different domain definition. Note that $N^F = N^{F7}$ is equivalent to $M = M^2$. Such domain difference arises from the switch of reference frame. Likewise, case (i) in Eq. (22) represents Scenario 1) and case (ii) represents Scenario 2) and 3).

We then directly apply Eqs.(20) and (22) to analyze the system commute cost. Given the number of parking space during morning peak $M$ , as shown in Figure 6, we have the following proposition.

**Proposition 4.2** *For given $M \le N^{C0}$, there is always an optimal number of eFHVs $N^F = N^{FO}(M)$, which leads the system commute cost to a minimum without any unused parking space.*

$$N^{FO}(M) = \begin{cases} N^{F8}, & \text{if } M > M^7; \\ N^{F7}, & \text{if } M \le M^7 \text{ and } M > M^6; \\ N^{F9}, & \text{if } M \le M^7 \text{ and } M \le M^6. \end{cases} \tag{23}$$

**Proof.** See Appendix A. 2. □

By switching the frame of reference, the results to maximize total commute become different compared to 4.1 . When the current supply of eFHVs is less than the optimum at current $M$ , the system commute cost increases. The insufficient eFHV supply directly results in the individual cost increase of transit riders along with ride sourcing passengers and can possibly lower the individual cost of private auto riders. On the contrary, when the current supply of eFHVs exceed the optimum, the oversupplied eFHVs attract private auto users to switch their commute mode, and hence lower the parking demand, given $W \le W_0$. In fact, the supply of eFHVs can be regarded as the aggregated number of eFHVs on road, independent of their types or owners. Inspired by traditional taxi industry, the number of eFHVs on road can be managed via various regulations, e.g., entry restrictions, fleet size control. Our work provides regulators a macroscopic tool of quantity management of ride sourcing vehicles during morning commute. Based on our results, regulators can further set an optimal number of eFHVs licenses and distribute them to various ride sourcing service providers, particularly when there's a sparse number of parking space.

#### 4.2.2 Numerical example

We take the values of parameters as in 4.1 in this example. The efficiency of the optimal eFHVs supply is also evaluated. Comparison and contrast are carried out between the benchmark case when $N^F = 0$ and the optimal case when $N^F = N^{FO}(M)$. We then define the percentage of cost reduction as

$$\eta = \frac{TC\left(N^F = 0\right) - TC\left(N^F = N^{FO}(M)\right)}{TC\left(N^F = 0\right)}. \tag{24}$$

The contour of relative efficiency $\eta$ is then shown in Figure 7. When the total number of commuters is $100,000$, and the number of parking space provided is $50,000$, by providing $18,466$ eFHVs for morning commute, we can then achieve the optimum with $\eta = 34.4\%$ reduction of system commute cost from $3.79 \times 10^6$ (SGD$) to $2.49 \times 10^6$ (SGD$).

## 5. Concluding remarks

In this paper, we explore the equilibrium morning commute traffic in a single-origin-single-destination bottleneck model with parking spaces restraints and examine the effectiveness of parking spaces management as a solution to morning commute problem.

Notably, other than private autos and transit, the emerging transportation mode, ride-sourcing service is taken into account in our model. With technological advances, the supply of ride-sourcing vehicles, i.e., number of eFHVs, can be substantially increased compared to that of traditional FHVs. Such increase inevitably contributes to morning traffic congestion and changes the parking demand at CBD areas. The through traffic characteristic without parking of eFHVs is explicitly examined in our work. We first draw a comparison of system performance between the case with and without ride-sourcing. Unlike previous parking research (Zhang et al., 2011; Yang et al., 2013), the number of road users no longer limits to the parking spaces $M$ in view of ride-sourcing and thereby the commute cost becomes different. In most cases with parking space paucity, the presence of eFHVs fills up the unexploited road capacity and therewith lower the system commute cost.

Moreover, we notice that the morning commute traffic management via parking spaces control is unproductive except in times of severe scarcity of parking space or public's disinclination towards ride-sourcing service. When ride-sourcing service is in general cost-competitive in contrast with private auto driving, there is always an optimal number of parking spaces corresponding to different eFHV quantities, leading to system commute cost minimum at a certain level of eFHVs supply, on the condition that all parking spaces are reserved. As a general policy, it is worth noting that, with consideration of ride-sourcing, the supply of parking spaces larger than optimum can bring about the increase of system commute cost as well as unnecessary spending on operational cost of parking space. In this case, the parking space provision should be reduced. One of the most manageable schemes to reduce parking supply with flexibility is to block some curbside parking in peak hour. This is consistent with the findings in some previous studies, e.g., Arnott et al. (2015). While Arnott et al. (2015) suggested the curbside parking elimination to alleviate congestion with less parking cruising, we propose the curbside parking control from the perspective of system commute cost minimum considering ride-sourcing. In addition to quantity control of parking space, we also examine the optimal number of eFHVs in morning commute with a given number of parking space so as to minimize the system commute cost. This result indeed provides the regulators a reference of aggregated number of eFHV licenses to be released when they are regulating the emerging ride sourcing industry.

Nevertheless, there are some unsolved problems. The cases when ride-sourcing cost is less attractive than that of driving have not been discussed in the section of optimal parking supply introduction. In those cases, it is possible that not all eFHVs are occupied during morning peak and the results can be different. Last but not least, we distinguish eFHVs from private autos by its demand for parking. In reality, some ridesharing vehicles providing ride-sourcing service may also need parking at CBD during morning peak. It is rational to incorporate them in future studies.

**Appendix A. 1**

We will derive the optimal number of parking spaces regarding different quantities of eFHVs in this appendix. As mentioned in Section 4, all parking spaces are now reserved, and private auto drivers will not drive to work without parking reservation.

According to Eqs.(19) and (20), when there once exists a period with zero flow at the bottleneck, namely when $N^{\mathrm{C}} < M^2$ as shown in Figure 4, the system commute cost (denoted by $TC^{\mathrm{i}}$) becomes

$$TC^{\mathrm{i}} = \left[ \frac{\beta\gamma}{(\beta+\gamma)s} N^{\mathrm{C}} + \alpha S_0 + F \right] \cdot N^{\mathrm{C}} + \left[ R + \theta\left(N - N^F - N^{\mathrm{C}}\right) \right] \cdot \left(N - N^C\right), \quad \text{(A1)}$$

where $N^F + N^{\mathrm{C}} < N$. The first-order derivative of $TC^{\mathrm{i}}$ with respect to $N^{\mathrm{C}}$ is then given by

$$\frac{dTC^{\mathrm{i}}}{dN^{\mathrm{C}}} = 2\left( \frac{\beta\gamma}{(\beta+\gamma)s} + \theta \right) \cdot N^{\mathrm{C}} + F - R + \alpha S_0 + \theta N^F - 2\theta N, \quad \text{(A2)}$$

When $N^{\mathrm{C}} > M^1$, $\frac{dTC^{\mathrm{i}}}{dN^{\mathrm{C}}} > 0$; when $N^{\mathrm{C}} \le M^1$, $\frac{dTC^{\mathrm{i}}}{dN^{\mathrm{C}}} \le 0$ .

Moreover, since $P^r \le P^f$ from (10),

$$\rightarrow \frac{\beta\gamma}{(\beta+\gamma)s} N^{\mathrm{C}} + \alpha S_0 + F \le R + \theta\left(N - N^F - N^{\mathrm{C}}\right),$$
$$\rightarrow N^{\mathrm{C}} \le -\frac{s(\beta+\gamma)\left(F - R + \alpha S_0 + \theta N^F - \theta N\right)}{\mathbb{C}} = M^4.$$

From $N^{\mathrm{C}} \ge 0$, there is $M^4 \ge 0$, $\rightarrow N^F \le \frac{-F + R - \alpha S_0 + \theta N}{\theta} = N^{\mathrm{F4}}$.

Besides, we have

$$\frac{\mathrm{d}^2 TC^{\mathrm{i}}}{\mathrm{d}\left(N^C\right)^2} = 2\left(\frac{\beta\gamma}{(\beta+\gamma)s} + \theta\right) > 0. \tag{A3}$$

As for the case when $N^{\mathrm{C}} \ge M^2$, i.e. there never exists a time window with zero flow at the bottleneck, the system commute cost (denoted by $TC^{\mathrm{ii}}$) is

$$\begin{aligned} TC^{\mathrm{ii}} = & \left\{F + (\alpha+\gamma)S_0 + \frac{\gamma[Ws + \beta\left(N^F + N^C\right) - \left(\theta\left(N - N^F - \mathrm{N}^{\mathrm{C}}\right) + R\right)s\Big]}{\beta s}\right\} \cdot N^{\mathrm{C}} \\ & + \left[R + \theta\left(N - N^F - N^{\mathrm{C}}\right)\right] \cdot \left(N - N^C\right). \end{aligned} \tag{A4}$$

The first-order derivative of $TC^{\mathrm{ii}}$ with respect to $N^{\mathrm{C}}$ is then given by

$$\begin{aligned} \frac{dTC^{\mathrm{ii}}}{dN^{\mathrm{C}}} = & \frac{2\mathbb{C}}{\beta s} N^C + F - R + \frac{\gamma N^F}{s} + \frac{\gamma(W - R)}{\beta} \\ & + (\alpha+\gamma)S_0 + \theta N^F - 2\theta N - \frac{\gamma\theta}{\beta}\left(N - N^F\right). \end{aligned} \tag{A5}$$

It can be verified that when $N^{\mathrm{C}} > M^3$, $\frac{dTC^{\mathrm{ii}}}{dN^{\mathrm{C}}} > 0$; when $N^{\mathrm{C}} \le M^3$, $\frac{dTC^{\mathrm{ii}}}{dN^{\mathrm{C}}} \le 0$.

Similarly, we have $P^r \le P^f$ ,

$$\rightarrow F+(\alpha+\gamma)S_0+\frac{\gamma[Ws+\beta\left(N^F+N^C\right)-\left(\theta\left(N-N^F-\mathrm{N}^{\mathrm{C}}\right)+R\right)s\right]}{\beta s}\le R+\theta\left(N-N^F-N^{\mathrm{C}}\right),$$
$$\rightarrow N^{\mathrm{C}}\le\frac{\theta N(\beta+\gamma)s-\left[\beta F+\gamma W+\beta(\alpha+\gamma)S_0-R(\beta+\gamma)\right]s}{\mathbb{C}}-N^F=M^5.$$

From $N^{\mathrm{C}}\ge 0$, there is $M^5\ge 0$, $\rightarrow N^F\le\frac{(\beta+\gamma)(\theta N+R)-\beta F-\beta(\alpha+\gamma)S_0-\gamma W}{\mathbb{C}}=N^{\mathrm{F5}}$. The second-order derivative of $TC^{\mathrm{ii}}$ is

$$\frac{\mathrm{d}^2TC^{\mathrm{ii}}}{\mathrm{d}\left(N^C\right)^2}=\frac{2\mathbb{C}}{\beta s}>0\,. \tag{A6}$$

Though Eqs. (A2)(A3)(A5)(A6) show that $TC^{\mathrm{i}}$ and $TC^{\mathrm{ii}}$ are strictly convex in $N^C$ respectively, the global total cost function $TC$ is a piecewise function and is not always convex in $N^C$.

Comparing $M^1$, $M^2$ and $M^3$, we have

$$\begin{aligned}&M^1\ge M^2,\ \text{iff}\ N^F\ge N^{\mathrm{F1}};\\&M^3\ge M^2,\ \text{iff}\ N^F\ge N^{\mathrm{F2}}.\end{aligned} \tag{A7}$$

Without considering the non-negativity of $N^{\mathrm{C}}$ and $P^r\le P^f$, the minimum of $TC$ corresponding to each segment is

$$\begin{array}{lll}\text{If } N^{\mathrm{C}}\le M^2: & \text{if } N^F\ge N^{\mathrm{F1}}, & TC_{min}=TC(N^{\mathrm{C}}=M^2);\\ & \text{else} & TC_{min}=TC(N^{\mathrm{C}}=M^1);\\ \text{else:} & \text{if } N^F\ge N^{\mathrm{F2}}, & TC_{min}=TC(N^{\mathrm{C}}=M^3);\\ & \text{else} & TC_{min}=TC(N^{\mathrm{C}}=M^2).\end{array} \tag{A8}$$

Evidently, the result in (A8) is not completed and should be further combined with the other constraints including $M^4$, $M^5$. Then comparing $M^2$, $M^4$ and $M^5$, it can be verified that

$$\frac{M^4 - M^2}{M^5 - M^2} = \frac{\beta + \gamma}{\beta}; \tag{A9}$$

and

$$M^4 > M^5 \geq M^2, \text{if } N^F \geq N^{F3}; M^4 < M^5 \leq M^2, \text{if } N^F \leq N^{F3}. \tag{A10}$$

Furthermore, the relationship between $N^{F1}$, $N^{F2}$ and $N^{F3}$ can also be found. Let $\bar{N} = \frac{\theta \mathbb{D} s + \beta(\mathbb{D} - \mathbb{E} + \theta N)}{2\beta\theta + \theta^2 s}$, we have

$$N^{F2} \geq N^{F1}, \quad \text{iff } N \geq \bar{N}; \tag{A11}$$

When $N^{F3} \leq N^{F1}$, $\bar{N} \leq 0$, thus $N \geq \bar{N}$ is always satisfied, $\rightarrow N^{F2} \geq N^{F1}$. Thus, it is impossible to have the cases when

$$N^{F3} \leq N^{F2} \leq N^{F1} \text{ or } N^{F2} \leq N^{F3} \leq N^{F1}. \tag{A12}$$

In addition, when $N < 2\bar{N}$, $N^{F4} < N^{F1}$; when $N < \frac{(2\beta + \gamma)(2\beta + \theta s)}{\beta(2\beta + \theta s + \gamma)} \bar{N}$, $N^{F4} < N^{F2}$; when $N < \left(2 + \frac{\theta s}{\beta}\right) \bar{N}$, $N^{F5} < N^{F3}$; thus when $N < \bar{N}$, namely when $N^{F2} < N^{F1}$, there is $N^{F4} < N^{F2} < N^{F1}$ and $N^{F5} < N^{F3}$.

Now from (A12), the optimal parking spaces problem can be separated into four different cases depending on the relationship of $N^{F1}$, $N^{F2}$ and $N^{F3}$, i.e., ①;②;③;④. For different $N^F$, the optimal number of parking spaces $M^O(N^F)$ can be calculated below in Table Appendix A.1.

**Table Appendix A.1 Optimal Number of Parking Spaces $M^O(N^F)$**

| $N^F$ | $\mathbb{S}=N^{F3}$, ① $\mathbb{M}=N^{F1}$, $\mathbb{L}=N^{F2}$; | $\mathbb{S}=N^{F1}$, ② $\mathbb{M}=N^{F3}$, $\mathbb{L}=N^{F2}$; | $\mathbb{S}=N^{F1}$, ③ $\mathbb{M}=N^{F2}$, $\mathbb{L}=N^{F3}$; | $\mathbb{S}=N^{F2}$, ④ $\mathbb{M}=N^{F1}$, $\mathbb{L}=N^{F3}$; |
|---|---|---|---|---|
| $(-\infty,\mathbb{S}]$ | $\min\{M^1,M^4\}$ | $\min\{M^1,M^4\}$ | $\min\{M^1,M^4\}$ | $\min\{M^1,M^4\}$ |
| $(\mathbb{S},\mathbb{M}]$ | $M^1$ | $M^4$ | $M^4$ | 0 |
| $(\mathbb{M},\mathbb{L}]$ | $M^2$ | $M^2$ | $M^4$ | 0 |
| $(\mathbb{L},+\infty)$ | $\min\{M^3,M^5\}$ | $\min\{M^3,M^5\}$ | $\min\{M^3,M^5\}$ | 0 |

Note:

1.For each case, $\mathbb{S}$, $\mathbb{M}$ and $\mathbb{L}$ represent the relationship of $N^{F1}$, $N^{F2}$ and $N^{F3}$, where $\mathbb{S}\le\mathbb{M}\le\mathbb{L}$, e.g., in case ① we have $N^{F3}\le N^{F1}\le N^{F2}$.

2. It is assumed that in this table $N^F$ satisfies that $N^F\le N^{F4}$ when $N^C<M^2$, or $N^F\le N^{F5}$ when $N^C\ge M^2$; otherwise $M^4\le 0$, $N^C\ge M^4$ when $N^C<M^2$, and $M^5\le 0$, $N^C\ge M^5$ when $N^C\ge M^2$, $P^r\ge P^f$, no one prefers private autos and the optimal number of parking spaces should be zero.

3. In case ④, $N^{F2}\le N^{F1}, \rightarrow N^{F4}\le N^{F2}$ as mentioned earlier. When $N^C<M^2$, if $N^F\ge N^{F2}$ $\rightarrow N^F\ge N^{F4}$, from Note 2., the optimal number of parking spaces should be zero. When $N^C\ge M^2$, as it can be verified that $M^3\le 0$ is satisfied when $N^{F2}\le N^{F1}$, $\rightarrow \min\{M^3,M^5\}\le 0$. Due to the non-negativity of $N^C$, the optimal number of parking spaces is zero.

After summarizing the results in Table Appendix A.1, the optimal number of parking spaces $M^O(N^F)$ is given as shown in Eq.(21).

**Appendix A. 2**

Following Appendix A.1, when $N^{\mathrm{C}} < M^2$, i.e., $N^F < N^{F7}$, it is Scenario 1). The system commute cost is shown in (A1). Moreover, the first-order derivative of $TC^{\mathrm{i}}$ with respect to $N^F$ is then given by

$$\frac{dTC^{\mathrm{i}}}{dN^F} = -\theta\left(N - N^C\right) < 0\,. \tag{A13}$$

Moreover, since $P^r \le P^f$ from (10),

$$\rightarrow \frac{\beta\gamma}{\left(\beta+\gamma\right)s} N^{\mathrm{C}} + \alpha S_0 + F \le R + \theta\left(N - N^F - N^{\mathrm{C}}\right),$$
$$\rightarrow N^F \le N^{F8}.$$

As for the case when $N^F \ge N^{F7}$, i.e., Scenario 2) and 3), the system commute cost is shown in (A4).The first-order derivative of $TC^{\mathrm{ii}}$ with respect to $N^F$ is then given by

$$\frac{dTC^{\mathrm{ii}}}{dN^F} = \frac{\gamma\left(\beta + s\theta\right)}{s\beta} N^{\mathrm{C}} - \theta\left(N - N^{\mathrm{C}}\right). \tag{A14}$$

It can be verified that when $N^C > M^6$, $\frac{dTC^{\mathrm{ii}}}{dN^{\mathrm{F}}} > 0$; when $N^{\mathrm{C}} \le M^6$, $\frac{dTC^{\mathrm{ii}}}{dN^{\mathrm{F}}} \le 0$.

Similarly, we have $P^r \le P^f$

$$\rightarrow F + \left(\alpha+\gamma\right)S_0 + \frac{\gamma[Ws + \beta\left(N^F + N^C\right) - \left(\theta\left(N - N^F - \mathrm{N}^{\mathrm{C}}\right) + R\right)s\,]}{\beta s} \le R + \theta\left(N - N^F - N^{\mathrm{C}}\right),$$
$$\rightarrow N^F \le N^{F9}.$$

One can further verify that

$$TC^{\mathrm{i}}\left(N^F = N^{F7}\right) = TC^{\mathrm{ii}}\left(N^F = N^{F7}\right). \tag{A15}$$

Therefore, given (A13)(A14)(A15), the global total cost function $TC$ is a continuous and convex function of $N^F$. When $N^C > M^6$, $TC$ reaches its minimum at $N^F = N^{F7}$, i.e., at the case when the first private auto happens to enter the bottleneck right after the last eFHV leaves the bottleneck; when $N^C < M^6$, $TC$ reaches its minimum at $N^F = N^{F0}$, i.e., at the case when all road users are eFHVs only.

However, the minimum is not stable by setting $N^F$ to $N^{F7}$ or $N^{F0}$ directly. The actual number of private auto users may differ when ride sourcing is more cost competitive than private auto driving, i.e., $P^r > P^f$. Besides, when $N^F = N^{F0}$, all parking spaces become idle.

Comparing $N^{F7}, N^{F8}$ and $N^{F9}$, we have

$$\frac{N^{F9} - N^{F7}}{N^{F8} - N^{F7}} = \frac{\beta\theta s}{\mathbb{C}} < 1, \tag{A16}$$

and

$$\begin{aligned} &N^{F8} \geq N^{F9} \geq N^{F7} \quad \text{if } M \leq M^7; \\ &N^{F8} \leq N^{F9} \leq N^{F7} \quad \text{if } M \geq M^7. \end{aligned} \tag{A17}$$

One can verify that when $M \leq N^{C0}$, $N^{F7}, N^{F9} > 0$, $N^{F8} \geq 0$.

When $M \geq M^7$, if it is in Scenario 2) or 3) at equilibrium, i.e., $N^F > N^{F7}$, we have $P^r > P^f$, the actual number of private auto users decreases until $M = M^5$, i.e., $N^F = N^{F9}$. Since $N^F = N^{F9} \leq N^{F7}$ when $P^r = P^f$, the equilibrium will go to Scenario 1). Once it is in Scenario 1), $N^{F8}$ bounds and finally $N^C = M^4$. Therefore, it is impossible for the equilibrium to stay at Scenario 2) or 3), when $M \geq M^7$. Namely no matter what $N^F$ is, when $M \geq M^7$, we have Scenario 1) at equilibrium. By setting $N^F = N^{F8}$, $TC$ reaches its minimum without unused parking space.

When $M < M^7$, it is possible for the equilibrium to stay in any scenarios. If $N^C > M^6$, as discussed, $TC$ reaches its minimum at $N^F = N^{F7} \leq N^{F9} \leq N^{F8}$. If $N^C \leq M^6$, $TC$ reaches its minimum at $N^F = N^{F9} \leq N^{F0}$. After summarizing the foregoing results, the optimal number of eFHVs $N^{FO}(M)$ is given as shown in Eq.(23).